\documentclass[12pt,preprint]{aastex}

\usepackage{epsfig}
\usepackage{amssymb} 
\usepackage{lscape}     

\newcommand{\msun}{\ensuremath{\, {M}_\odot}}
\newcommand{\ct}{\ensuremath{^{13}\mem{C}}}
\newcommand{\cd}{\ensuremath{^{12}\mem{C}}}

\newcommand{\nfo}{\ensuremath{^{14}\mem{N}}}
\newcommand{\nfi}{\ensuremath{^{15}\mem{N}}}
\newcommand{\osi}{\ensuremath{^{16}\mem{O}}}

\newcommand{\oei}{\ensuremath{^{18}\mem{O}}}

\newcommand{\fni}{\ensuremath{^{19}\mem{F}}}

\newcommand{\mem}[1]{\ensuremath{\mathrm{ #1}}}
\newcommand{\spr}{\mbox{$s$ process}}

\newcommand{\fig}[1]{Figure\,\ref{#1}}
\newcommand{\secr}[1]{\S\,\ref{#1}}

\newcommand{\tab}[1]{Table\,\ref{#1}}

\slugcomment{Submitted to ApJ \today}

\shorttitle{Production of \fni\ in AGB stars}

\received{2002 November 26}
\begin{document}

\title{Reaction Rates Uncertainties and the Production \\ of \fni\ in AGB Stars}
\author{Maria Lugaro}
\affil{Institute of Astronomy, University of Cambridge, Madingley Road, \\ Cambridge CB3
0HA, United Kingdom }
\email{mal@ast.cam.ac.uk}
\author{Claudio Ugalde}
\affil{Department of Physics, University of Notre Dame, Notre Dame, IN 46556}
\email{ugalde.1@nd.edu}
\author{Amanda I. Karakas}
\affil{Institute for Computational Astrophysics,
Department of Astronomy \& Physics, Saint Mary's University, Halifax, Canada}
\email{akarakas@ap.stmarys.ca}
\author{Joachim G\"orres and Michael Wiescher}
\affil{Department of Physics, University of Notre Dame, Notre Dame, IN 46556}
\email{wiescher.1@nd.edu, goerres.1@nd.edu}
\author{John C. Lattanzio}
\affil{School of Mathematical Sciences, PO Box 28M, Monash University, \\ Victoria
3800 Australia}
\email{j.lattanzio@sci.monash.edu.au}
\author{Robert C. Cannon}
\affil{Institute of Adaptive and Neural Computation, Division of Informatics, \\ 5 Forrest
Hill, Edinburgh EH1 2QL, United Kingdom}
\email{robert.cannon@ed.ac.uk}

\begin{abstract}

We present nucleosynthesis calculations and the resulting \fni\ stellar yields
for a large set of models with different masses and metallicity. During the
Asymptotic Giant Branch (AGB) phase \fni\ is produced as a consequence of
nucleosynthesis occurring during the convective thermal pulses and also during the
interpulse periods if protons from the envelope are partially mixed in the top
layers of the He intershell (partial mixing zone). We find that the production
of fluorine depends on the temperature of the convective pulses, the amount of 
primary \cd\ mixed into the envelope by third dredge up and the extent of the 
partial mixing zone. Then we perform a detailed analysis of the reaction rates 
involved in the production of \fni\ and the effects of their uncertainties. We 
find that the major uncertainties are associated with the 
$^{14}$C($\alpha,\gamma$)\oei\ and the \fni($\alpha,p$)$^{22}$Ne reaction rates. 
For these two reactions we present new estimates of the rates and their 
uncertainties. In both cases the revised rates are lower than previous estimates. 
The effect of the inclusion of the partial mixing zone on the production of 
fluorine strongly depends on the very uncertain $^{14}$C($\alpha,\gamma$)\oei\ 
reaction rate. The importance of the partial mixing zone is reduced when 
using our estimate for this rate. Overall, rate uncertainties result in 
uncertainties in the fluorine production of about 50\% in stellar models with 
mass $\simeq$ 3 \msun\ and of about a 
factor of 5 in stellar models of mass $\simeq$ 5 \msun. This larger effect at 
high masses is due to the high 
uncertainties of the \fni($\alpha,p$)$^{22}$Ne reaction rate. 
Taking into account both the uncertainties related to the partial mixing zone 
and those related to nuclear reactions, the highest values of \fni\ 
enhancements observed in AGB stars are not matched by the models. This is a 
problem that will have to be revised by providing a better understanding of the 
formation and nucleosynthesis in the partial mixing zone, also in relation to 
reducing the uncertainties of the $^{14}$C($\alpha,\gamma$)$^{18}$O reaction 
rate. At the same time the possible effect of Cool Bottom Processing at 
the base of the convective envelope should be included in the computation of AGB 
nucleosynthesis. This process could in principle help matching the 
highest \fni\ abundances observed by decreasing the C/O ratio at the surface of 
the star, while leaving the \fni\ abundance unchanged.

\end{abstract}

\keywords{stars: AGB
--- nuclear reactions, nucleosynthesis, abundances, fluorine}

\section{Introduction}
\label{sec:intro}

Spectroscopic observations show that in giant stars of type K, M,
MS, S, SC and C the fluorine abundance is enhanced by factors of 
2 to 30 with respect to the solar abundance \citep*{jorissen:92}. 
These low-mass stars are the only astrophysical site 
observationally confirmed to produce fluorine. Hence they 
are good candidates to account for the Galactic abundance of this 
element, even though recent observations of \fni\ in the LMC and 
$\omega$ Cen, where the abundance ratio of F/O declines with the oxygen 
abundance, may support the hypothesis that most fluorine is 
produced instead by massive stars \citep{cunha:03,renda:04}. 
In any case the fluorine abundances observed in giant stars
are of considerable importance in constraining the properties of
Asymptotic Giant Branch (AGB) models. In AGB stars H- and 
He-shell burning with subsequent He-pulse driven convection 
(thermal pulse) change
the abundance distribution between the H- and the He-burning
shells (He intershell). Partial He burning in the He intershell
converts He into \cd. 
After the occurrence of a thermal pulse, the convective envelope can
penetrate the He intershell and dredge up material to the surface (third 
dredge up, TDU). The stellar atmosphere
becomes progressively rich in carbon, thus explaining the observed
sequence of carbon enrichment from M to S and C stars. These stars
also show enhancements of elements produced by slow neutron
captures (\spr) and are believed to be the main site for the
production of $s$-process nuclei with mass above
$\simeq$ 90 
\citep{gallino:98,travaglio:99,goriely:00,travaglio:01}.

The observed enhancements of fluorine in AGB stars indicate a
positive correlation with the carbon enhancements. This can be explained
if \fni\ is also produced in the He intershell and then dredged up
to the surface together with \cd\ and $s$-process elements.
\citet{jorissen:92} proposed the following nucleosynthesis path
for the production of \fni\ in the He intershell of AGB
stars. Neutrons produced via the \ct($\alpha,n$)$^{16}$O
reaction can be captured by \nfo\, which is enriched from the preceding 
H-burning stage where the CNO cycle dominates.
The reaction \nfo($n,p$)$^{14}$C has a high cross section and
produces free protons and $^{14}$C which is converted
by $\alpha$-capture to $^{18}$O; alternatively $^{18}$O can also
be produced by $\alpha$-capture on \nfo\ with subsequent
$\beta$-decay of $^{18}$F. In core He burning $^{18}$O is converted by 
further $\alpha$-capture to produce $^{22}$Ne via the reaction 
$^{18}$O($\alpha,\gamma$)$^{22}$Ne, however, in the He intershell
$^{18}$O and protons are present at the same time triggering the
alternative reaction path $^{18}$O($p,\alpha$)$^{15}$N. Subsequent
$\alpha$-capture on $^{15}$N eventually leads to the production of
\fni, via $^{15}$N($\alpha, \gamma$)$^{19}$F. The 
$^{15}$N($p,\alpha$)$^{12}$C reaction competes with the 
$^{18}$O($p,\alpha$)$^{15}$N reaction and removes both protons and 
$^{15}$N from the chain of production of $^{19}$F.
The abundance of \fni\ is determined by the reaction rates associated with this
rather complex production path and by the \fni\
destruction reaction in the He intershell, \fni($\alpha,p$)$^{22}$Ne.

In summary, the reactions that contribute to or affect the
production of fluorine are: $$ ^{13}{\rm C}(\alpha,n)^{16}{\rm O},
^{14}{\rm N}(n,p)^{14}{\rm C}, ^{14}{\rm N}(\alpha,
\gamma)^{18}{\rm F}, ^{14}{\rm C}(\alpha,\gamma)^{18}{\rm O},
^{18}{\rm O}(\alpha,\gamma)^{22}{\rm Ne}, 
^{15}{\rm N}(p,\alpha)^{12}{\rm C},
$$ together with the
alternative reaction chain $$ ^{18}{\rm O}(p,\alpha)^{15}{\rm
N}(\alpha, \gamma)^{19}{\rm F}(\alpha,p)^{22}{\rm Ne}. $$ 

The theoretical studies of \citet{forestini:92} and 
\citet*{mowlavi:96} found that the above described chain is 
activated in the convective pulse when neutrons are released 
by \ct\ from the H-burning ashes. However, only the lowest observed 
abundances of \fni\ could be explained. An extra 
amount of \ct\ is required to produce the observed \fni, and also to 
match the observed enhancements of \spr\ elements. 
At the end of each
TDU where the convective envelope expands into the stable
radiative intershell zone extra-mixing processes could lead to 
the formation of a zone where protons and \cd\ are partially 
mixed (partial mixing zone). This would lead to additional 
production of
\ct\ by the \cd($p,\gamma$) reaction in the top layers of the He
intershell. Models including hydrodynamical overshoot
\citep{herwig:00}, rotation \citep{langer:99} or the effect of
gravity waves \citep{denissenkov:03} have in fact produced a 
partial mixing zone resulting in the formation of a \ct\ {\it
pocket}. \citet{straniero:95} showed that the \ct\ formed in the
pocket is completely destroyed by the \ct($\alpha,n$)$^{16}$O
reaction before the onset of the next convective pulse. By means
of a parametric representation of the partial mixing zone
\citet{gallino:98} and \citet{goriely:00} showed that this model
can explain the observed properties of the \spr\ in AGB stars. In the \ct\ 
pocket $^{15}$N is produced at conditions where the value 
of the proton to \cd\ ratio is close to unity
\citep*[see also][]{mowlavi:98}. This $^{15}$N is converted into
$^{19}$F when the pocket is ingested in the following convective
pulse. \citet{goriely:00} analyzed the effect of the presence of
the partial mixing zone on the nucleosynthesis of fluorine. 
These authors concluded that also by taking into account the 
nucleosynthesis in the partial mixing zone only the less
fluorine-enriched stars could be explained. The possible effect on
the nucleosynthesis in the partial mixing zone due to stellar
rotation also did not seem to improve the match with observations
\citep*{herwig:03}. 

The aims of this paper are to update the study of the production of \fni\ 
in AGB stars and to explore the impact of the uncertainties of nuclear 
reaction
rates on the abundance of fluorine produced in the framework of the current AGB 
star models. First we introduce the production of \fni\ in AGB models 
of a large range of masses and 
metallicities. We calculate the stellar structure and then follow the 
nucleosynthesis by making use of a postprocessing code.
Our computations represent an improvement with respect to previous 
computations for several reasons. First, we find the TDU to occur 
self-consistently after a certain number of thermal pulses, hence we do 
not parametrize this process as done in all the previous studies.
If it is true that the amount of TDU is still uncertain \citep[see 
e.g.][]{frost:96,mowlavi:99} and hence can be parametrized, our approach
is more consistent in the fact  
we not only deal with TDU as a way of mixing fluorine to the stellar 
surface but also take into account the feedback effect of TDU on the 
nucleosynthesis of \fni\ in the He intershell. As we will show, this 
feedback has a large impact on the production of \fni. 
Secondly, our postprocessing code follows the nucleosynthesis throughout 
all the different thermal pulses previously computed by the evolutionary 
code. This was done by \citet{mowlavi:96} for three stellar  
models with a limited number of pulses, but without including a partial 
mixing zone. \citet{goriely:00} included a partial mixing zone in their 
calculations but only followed the nucleosynthesis ``during one 
representative interpulse and pulse phase'' hence missing the 
possible effects due to variations of the thermodynamic features 
of each thermal pulse. 
Finally, our postprocessing code computes abundances of nuclei up to iron 
solving simultaneously the changes due to nuclear reactions and those due
to mixing, when convection is present. This allows us, for example, to 
properly model the nucleosynthesis that occurs at the delicate moment 
when the H-burning ashes are progressively ingested in the convective 
pulse and the \ct\ present in the ashes burns via the ($\alpha,n)$ 
reaction while the ingestion is occurring.

We discuss and compare results from a large set of stellar models, analyze 
in detail the impact of the introduction of the partial mixing zone and of 
the reaction rate uncertainties on the 3 \msun\ $Z=0.02$ model and then 
present upper and lower limits for the production of fluorine in several 
selected models. On top of the comparison with spectroscopic observations of 
AGB stars, our results are of relevance when studying the 
Galactic chemical evolution of fluorine, as done recently by 
\citet{renda:04}.
The evolutionary and nucleosynthesis codes are presented in 
\S2. The production of fluorine in a large range of stellar models is 
discussed in \S3. The effect of introducing a partial mixing zone is discussed 
in \S4. The nuclear 
reactions contributing to the production of \fni\ are discussed in \S5 
together with the effect of their uncertainties on the production of \fni. In 
\S6 we present a final discussion and possible directions for future work.

\section{Evolutionary and nucleosynthesis codes}
\label{sec:codes}

We computed the stellar structure for a large range of masses (from $M$=1 
to 6.5 \msun) and metallicities ($Z$=0.0001, 0.004, 0.008 and 0.02) 
starting from the zero-age main sequence up throughout many thermal pulses 
during the AGB phase using the Mount Stromlo Stellar Structure Program 
\citep{wood:81,frost:96}. Mass loss is modelled on the AGB phase
following the prescription of \citet{vassiliadis:93}, which
accounts for a final {\it superwind} phase. Using the prescription for 
unstable convective/radiative boundaries described in detail by 
\citet{lattanzio:86} we find the third 
dredge up to occur self-consistently for masses above 
2.25 \msun\ at $Z$=0.02, above 1.5 \msun\ at $Z$=0.008, above 
1.25 \msun\ at $Z$=0.004 and for all the computed masses at $Z$=0.0001.
More details regarding these calculations can be found 
in \citet{karakas:03} and for the 3 \msun\ $Z$=0.02 model
in \citet{lugaro:03}.

To calculate the nucleosynthesis in detail we have used a
postprocessing code that calculates abundance changes due to
convective mixing and nuclear reactions \citep{cannon:93}.
The stellar structure inputs, such as
temperature, density, extent of convective zones, mixing length
and mixing velocity as functions of mass and model number, are
taken from the stellar evolutionary computations. 
Between evolution models the 
postprocessing code creates its 
own mass mesh, resolving regions undergoing rapid changes in composition
and using a combination of Lagrangian and non-Lagrangian points.
Convective mixing is done time dependently, with no assumptions of
instantaneous mixing. To model this, a ``donor cell'' scheme is
adopted in which each nuclear species is stored as two variables
representing two streams, one moving upward and one moving
downward. At each mass shell matter flows freely from above or
below with a certain degree of mixing, and is also exchanged
between adjacent cells, from one stream to the other. 

Our nucleosynthesis network is based on 74 nuclear species, 59
nuclei from neutrons and protons up to sulphur and with another 14
nuclei near the iron group to allow neutron capture on iron seeds.
There is also an additional ``particle'' $g$ for counting the
number of neutron captures occurring beyond $^{61}$Ni, which  
simulates the $s$ process as neutron $sink$. The initial 
abundances in the postprocessing calculations are taken from 
\citet{anders:89}. All proton, $\alpha$, neutron captures and $\beta$ 
decays involving
the species listed above are included in the nuclear network
summing up to 506 reactions. The bulk of reaction rates are from
the REACLIB Data Tables of nuclear reaction rates based on the
1991 updated version of the compilation by \citet*{thielemann:86}.
The reaction rate table has been updated using the latest
experimental results, which are listed in Appendix A. The reaction
network is terminated by a neutron capture on $^{61}$Ni followed
by an {\it ad hoc} decay with $\lambda$ = 1 s$^{-1}$ producing the
particle represented by the symbol $g$:
$^{61}$Ni($n,\gamma$)$^{62}$Ni $\rightarrow^{61}$Ni + $g$.
Following the method of \citet{jorissen:89} neutron captures on
the missing nuclides are modelled by neutron $sinks$, meaning that
the $^{34}$S($n,\gamma$)$^{35}$S and the
$^{61}$Ni($n,\gamma$)$^{62}$Ni reactions are given some averaged
cross section values in order to represent all nuclei from
$^{34}$S to $^{55}$Mn and from $^{61}$Ni to $^{209}$Bi,
respectively \citep[see also][]{lugaro:03,herwig:03}.

\section{Results for the production of fluorine}
\label{sec:allmodels}

Our model predictions for the final \fni\ intershell abundance are shown in
\fig{fig:intershell}. Note that these calculations do not include a partial 
mixing zone. We find that the abundance of \fni\ in the intershell is mostly 
dependent on two model features. The first is the temperature at the base of 
the convective pulse. As discussed by \citet{mowlavi:96}, this temperature  
determines the efficiencies of the rates of production and destruction of \fni. 
Below $\simeq$ 2.2 $\times 10^8$ K \nfi\ is not efficiently converted into \fni, 
while above $\simeq$ 2.6 $\times 10^8$ K \fni\ starts being destroyed by the 
\fni($\alpha,p)^{22}$Ne reaction. The stellar model of 3 \msun\ $Z$=0.02 of 
\citet{mowlavi:96} was shown to have pulse temperatures around the above range 
and hence to be the most efficient case for the production of fluorine with 
respect to the other two models presented by these authors: a 3 \msun\ 
$Z$=0.001 and a 6 \msun\ $Z$=0.02 stars. In the latter case 
proton captures at the hot base of the convective envelope (hot bottom 
burning) contribute to the destruction of fluorine.
Also in our models the maximum abundance of \fni\ in the He intershell at the 
end of the computed evolution is observed to occur at around 3 \msun, even though the 
temperatures are higher in our models, up to $\simeq$ 3 $\times 10^8$ K.

The second parameter that determines the abundance of \fni\ in the intershell 
is the amount of TDU. This is demonstrated by the fact that the maximum \fni\ 
intershell abundance as a function of the stellar mass is about double in the 
case of $Z$=0.008 than in the case of $Z$=0.02, which could appear at first 
surprising. In fact one would expect to find a lower \fni\ abundance at 
$Z$=0.008 because the temperature in the convective pulse is slightly higher: in 
the $Z$=0.02 case it ranges from $2.52 \times 10^8$ K in the 10$^{\rm th}$ pulse 
to $3.05 \times 10^8$ K in the last pulse, while in the $Z$=0.008 
case the temperature is around $3 \times 10^8$ K in the last ten pulses. 
Moreover, one would expect to find the \fni\ abundance decreasing with the 
metallicity of the star since, when no partial mixing zone is included, its 
production depends on the amount of \ct\ in the H-ashes which is of secondary 
nature, i.e. depends on the CNO abundances in the star. However, the abundance of
\cd\ in the envelope is a function of the amount of TDU. Since in our $Z$=0.008
models the total mass dredged up by TDU is about twice that in the $Z$=0.02 
models, there is a strong effect on the production of \fni\ due to the 
primary contribution to \ct\ 
in the H-burning ashes coming from the dredged-up \cd. 
Also the reason why the abundance of \fni\
decreases for masses lower than about 3 \msun\ is mostly due to the lower TDU 
rather than to the lower temperature in the convective pulse. This is 
demonstrated by 
the fact that the abundance of \nfi\ in all cases is insignificant with 
respect to that of \fni, which means that the fraction of \nfi\ that has not 
burned into \fni\ is unimportant.
Out of all the models, a maximum value of 2.5 $\times 10^{-6}$ for the
final \nfi\ intershell mass fraction is computed for the 1 \msun\ $Z=$0.02 star,
compared to the final \fni\ mass fraction of 7 $\times 10^{-6}$.

When comparing with the previous results of \citet{mowlavi:96} we find major 
differences due to two main reasons. The first is the fact that we have 
computed a much larger number of thermal pulses than \citet{mowlavi:96}. 
For example, for the stellar model of 6 \msun\ $Z$=0.02 we have computed 
38 thermal pulses while \citet{mowlavi:96} computed 11 thermal pulses. 
Hence the temperature at the base of the last convective pulse, which increases 
with pulse number in AGB models, is higher  
in our calculations. In our 6 \msun\ $Z$=0.02 model the 
temperature reaches 3.5 $\times 10^8$ K in the last thermal pulse 
which is higher than the value found by \citet{mowlavi:96} of 2.8 $\times 
10^8$ K simply because our last pulse represents a more advanced phase 
of the evolution. Hence, our final \fni\ abundance in the He intershell for 
this case is more than an order of magnitude lower than that calculated by 
\citet{mowlavi:96}. On the other hand, because the TDU is self-consistently 
included in our calculations, we take into account the effect of the 
presence of primary \cd\ in the envelope discussed above and thus 
the final \fni\ abundance for the 3 \msun\ $Z$=0.02 case in our 
calculation is about double that presented by \citet{mowlavi:96}.
The same conclusion can be drawn when our results are compared with those of 
\citet{forestini:97}, which are very similar to the results from 
\citet{mowlavi:96}. 

The production of fluorine in AGB stars is of interest also in the light of the 
Galactic chemical evolution. In \fig{fig:yields} and \tab{tab:tab1} we 
present 
yields for \fni\ calculated for the different model shown in 
\fig{fig:intershell}. Yields are a direct function of the amount of TDU. 
They are calculated as net yields: $M = \int_0^{\tau} (X - 
X_0) \frac{dM}{dt} dt$ where $\tau$ is the total lifetime of the star, $dM/dt$ is 
the mass-loss rate and $X$ and $X_0$ refer to the current and initial 
mass fraction of \fni. The yield is positive if \fni\ is produced and negative 
if it is destroyed. 
The \nfi\ yields are typically negative, decreasing 
from $\sim$ 0 for stars of 1 \msun\ to $\simeq - 2 \times 10^{-5}$ for 
stars of 6 \msun. This means that this isotope is 
destroyed in all the models, except those with $Z$=0.0001 and mass 
higher than 2.25 \msun. The \nfi\ yield reaches a positive maximum of $4 \times 
10^{-6}$ for the highest mass model computed as this metallicity (5 \msun). 
This is due to a combination of different factors: (i) the temperature at the 
base 
of the convective envelope is as high as 9.7 $\times 10^7$ K in this model, at 
which temperature the \nfo($p,\gamma)^{15}$O reaction becomes as important as the 
\nfi($p,\alpha$)\cd\ reaction and \nfi\ can actually be produced by proton 
captures during hot bottom burning, (ii) the abundance of 
\nfo\ is extremely high because of the operation of strong TDU and hot bottom 
burning, and (iii) the initial \nfi\ abundance ($X_0$ in the formula 
above) is very low. The initial \fni\ abundance is also very small and hence the \fni\ 
yields for this metallicity are less negative for masses above 4 \msun\ compared to 
more metal-rich models of the same mass.

In \fig{fig:f19models} we compare some selected model predictions with the 
observations by \citet{jorissen:92}. The metallicity of the observed stars 
ranges from about $Z$=0.006 to about $Z$=0.04 with an average of 0.016. Hence 
the 2.5 \msun\ $Z$=0.004 model has a metallicity too low to be considered to 
match the observations and it is included in the figure only to illustrate the 
trend of our results with metallicity. The 3 \msun\ $Z$=0.008 model, which has 
the highest final \fni\ abundance in the intershell, does not represent a good 
match to the stellar data. This is because the final C/O abundance in this 
model is 5.6, while the stellar data have C/O up to about 1.5. It follows that 
since the large \fni\ abundance in this model is a consequence of the
large \cd\ abundance in the envelope, we cannot take this model 
to explain the highest observed values. (We note though that stars with the 
high C/O ratio and high \fni\ abundance produced by this model may in 
principle exist but be obscured by their dusty envelopes). 
It should also be considered that the observational
data regarding SC stars require revision. For these stars it
is difficult to derive reliable abundances because of the poor
modeling of the atmospheres when C/O $\sim$ 1. 

The problem of matching the highest observed \fni\ abundance could 
be overtaken by the inclusion of extra-mixing processes at the base of the 
convective envelope, also referred to as Cool Bottom Processing. 
This process occurs during the first red giant phase in stars with 
$M \le 2.5$ \msun\ \citep[see e.g.][]{charbonnel:95}, possibly also 
during the AGB phase \citep*{nollett:03}, and results in lower \cd/\ct\ 
ratio than the standard models, as required by the observations.
This type of extra mixing is described as the circulation of   
material from the base of the convective envelope into the thin radiative region 
located on top of the H-burning shell. Here the material is processed by 
proton captures and then carried back to the envelope, thus producing the 
signature of CNO processing at the stellar surface.
Some of the MS, S stars with the highest [\fni/\osi] ratios for a
given C/O ratio are also enhanced in N, up to 2.5 times the initial value 
\citep[see \fig{fig:f19models} and discussion in][]{jorissen:92}. This 
N-enhancement could be due to Cool Bottom Processing. If 
this process is at work the surface \cd/\osi\ ratio would appear to 
be lower than computed in our calculations. On the other hand, if the 
temperature at which the material is carried by Cool Bottom Processing is 
lower than about 30 million degrees, at which value the \fni($p,\alpha$)\osi\ 
reaction is activated, then the \fni\ abundance would be unchanged. This is 
because the \fni\ production depends on the amount of \ct\ in the H-burning 
ashes which is a byproduct of CNO cycling, and would not in principle be 
different if the CNO cycling occurs only in the H-burning shell or also at 
the base of the convective envelope via Cool Bottom Processing. Then the 
theoretical curves of \fig{fig:f19models} would be simply shifted to the left 
making it easier to explain the observed \fni\ abundances, together with the N 
excess. Note that WZ Cas is the only Li-rich star of the sample and has very low 
\cd/\ct\ ratio, a composition that is in agreement with this extra mixing.
Cool Bottom Processing in the AGB phase is very uncertain and detailed 
computations are not available yet. Since it has has not been included in our 
computations we cannot draw any quantitative conclusions on its possible 
effects.

Limiting the discussion to our current models, as shown in 
\fig{fig:f19models}, at C/O $\sim$ 1 the 3 \msun\ $Z$=0.02 
model shows a higher \fni\ abundance in the envelope than the 3 
\msun\ $Z$=0.008 model. In the 5 \msun\ $Z$=0.02 model, hot bottom burning is 
at work 
hence both \cd\ and \fni\ are destroyed.
When comparing to previous calculations for the 3 \msun\ $Z$=0.02 model 
we find that our final [\fni/\osi] ratio in the envelope is about 0.25 dex 
higher than that computed by \citet{forestini:97} and \citet{mowlavi:96} 
for the same C/O $\simeq$ 1.2 ratio, which reflects our higher fluorine 
intershell abundance.

\subsection{The impact of the partial mixing zone}
\label{sec:3msun}

To study the effect of the introduction of a partial mixing zone 
we have included artificially in the postprocessing calculation 
a partial mixing zone at the end of each TDU episode. We have made the choice 
to include the partial mixing zone only when TDU occurs because during TDU
a sharp discontinuity is produced between the convective envelope and the radiative 
intershell, which is a favourable condition for the occurrence of mixing 
\citep[see e.g.][]{iben:82}. Since the question of the specific shape of the H-profile 
and the mixing processes leading to the partial mixing zone is still open we
opted for a reasonable choice of the proton profile in which the
number of protons decrease exponentially with the mass depth below
the base of the convective envelope. We define as the partial mixing
zone the region where the number of protons range from the
envelope value to $X_p=10^{-4}$. In this way about 1/4 of the 
extent of the partial mixing zone has a number of protons between 
$X_p=$0.002 and $X_p=$0.02, corresponding to the efficient range for 
the production of \nfi\ \citep[see][]{goriely:00}. Note that 
\citet{goriely:00} defined the partial mixing zone with the number of 
protons ranging from the envelope value to $X_p=10^{-6}$ so that 
$\sim$1/6 of its extent corresponds to the efficient range for the 
production of \nfi. For the extent of the partial mixing zone we 
considered a value of $M_{pmz}$= 0.001 \msun\, i.e. 1/15 of the 
mass of the last convective pulse for the 3 \msun\ $Z$=0.02 model. The 
dilution is higher for earlier pulses which have higher mass. This is a 
typical value adopted 
in the previous nucleosynthesis calculations \citep{gallino:98,goriely:00}.

In \fig{fig:m3z02pulses} we show the abundance of \nfi\ and \fni\ in the
intershell during the period of convective instability following each thermal pulse 
for the 3 \msun\ $Z$=0.02 model. The final abundances in  
each pulse can be identified as those corresponding to the pulse number 
tick-mark in the x-axis. At the beginning of a thermal pulse, while the 
convective instability is ingesting the H-burning ashes, \nfi\ is 
produced and its abundance sharply increases. At the same time the 
abundance of \fni\ decreases because of the dilution of the intershell 
material with H-burning ashes where the abundance of \fni\ is
solar. Subsequently, \nfi\ is transformed into \fni. In thermal pulses
followed by TDU in our model, i.e. from the 10$^{\rm th}$ thermal pulse onward, 
almost all \nfi\ is destroyed. The maximum temperature at the base of the 
10$^{\rm th}$ thermal pulse is $= 2.52 \times 10^8$ K and \nfi\ is reduced 
to about 1/10$^{\rm th}$ of its initial abundance in this pulse. In 
later pulses the temperature grows reaching $3.05 \times 10^8$ K in 
the last thermal pulse so that \nfi\ is destroyed with even higher efficiency. 
In the very last few pulses also about 25\% of the \fni\ produced is 
destroyed. The effect of the partial mixing zone appears after the
11$^{\rm th}$ thermal pulse where we observe large changes to the
intershell abundances. For example, the amount of \nfi\ and \fni\ suddenly
increase: in the 11$^{\rm th}$ thermal pulse the abundance of \fni\ is about 
2.5 times higher than that in the 10$^{\rm th}$ thermal pulse.
The final abundance of \fni\ in the intershell is $\simeq$ 70\% higher with respect 
to the case with no partial mixing zone included (shown in \fig{fig:intershell}). 

The extent in mass and the proton 
profile of the partial mixing zone are very uncertain parameters. 
Most studies that have self-consistently produced a partially mixed 
zone find that the extent in mass is smaller than the 0.001 
\msun\ value that we have used. The 
computed $M_{pmz}$ is of the order of 10$^{-6}$ \msun\ with rotation, of 
10$^{-5}$ \msun\ with overshoot (but depending on the free overshoot 
parameter!) and of 10$^{-4}$ \msun\ with gravitational waves. A partial 
mixing zone of larger extent, 5 $\times$ 10$^{-4}$ \msun, was reported to 
result from semiconvection in a low-metallicity star \citep{hollowell:88}. 
On the other hand previous nucleosynthesis studies have artificially 
considered partial mixing zones of extent up to 1/10 of the mass of the 
convective pulse. To check the uncertainty introduced by the extent of 
the partial mixing zone we varied this parameter 
thus computing three cases in total: one without zone included, and the other two 
with the mass of the zone equal to $M_{pmz}$= 0.001 \msun\ and 0.002 \msun.

The results are presented in \fig{fig:f19pocket} and show that the variation of 
the final abundance of \fni\ in the envelope is up to a factor of $\sim$2 when the 
mass extent of the partial mixing zone is varied in the range described above.
This could probably be considered as an upper limit for the uncertainty since a 
mass of $M_{pmz}$= 0.002 \msun\ is a large value to consider within the 
framework of the current models. A higher mass in fact would imply that the mixing 
process carrying protons into the He intershell region involves a large 
fraction of the intershell mass, which is not what the current studies 
indicate. We can only make a qualitative comparison with the results obtained by 
\citet{goriely:00} since the stellar model considered is different as 
well as the computation procedure. Our case with $M_{pmz}$= 0.001 \msun\ and the 
case presented by 
\citet{goriely:00} with $\lambda_{pm}=M_{pmz}$/$M_{convective\, shell}$=0.1 should 
have a very similar value for the extent of the 
region where the production of 
\nfi\ is efficient in the partial mixing zone, corresponding to $\sim$ 1/60 of 
the total mass of the intershell. However, for this case the increase 
in the [\fni/\osi] ratio that we computed is more than 0.3 dex higher for the same 
C/O value around 1.2, than that presented in Figure 12 of \citet{goriely:00}. 
This is probably due to the fact that we have self-consistently 
taken the TDU into account. 

The introduction of a partial mixing zone in some 
selected stellar models is illustrated in \tab{tab:tab2}, where the \fni\ 
yields are reported from computations performed without (Column 2) and with 
(Column 3) the inclusion of the partial mixing zone. Since we have introduced 
the same partial mixing zone in all the models, and since in the 5 \msun\ 
model the extent in mass of the 
intershell is about 0.005 \msun, half the value than all the other models, 
the dilution factor of the pocket material in this case is a factor of two lower. 
Note also that in principle we do not know if and how the formation of the 
partial mixing zone is a function of the stellar properties.
In the stellar models 
with mass $\simeq$ 3 \msun\ the effect of the partial mixing zone introduces 
a factor of 2.6 uncertainty in the final yield, in the 5 \msun\ 
$Z=$0.02 model the uncertainty is of about a factor of 4, while in the 
low mass model, 1.75 \msun, the uncertainty is of a factor of 14 
in the final yield. However, as will be discussed in \secr{sec:summa}, 
this effect strongly depends on the uncertainties associated with the 
$^{14}$C($\alpha,\gamma$)\oei\ reaction rate.

\section{Summary of reaction rate studies}
\label{sec:summa}

There has been a considerable effort and improvement in the
determination of the nuclear reaction rates over the last few
years since the early $^{19}$F nucleosynthesis studies.  In
particular new measurements of key reactions such as
$^{14}$C($\alpha,\gamma$)$^{18}$O,
$^{14}$N($\alpha,\gamma$)$^{18}$F,
$^{15}$N($\alpha,\gamma$)$^{19}$F,
$^{18}$O($\alpha,\gamma$)$^{22}$Ne provided new information on low
energy resonances which were ignored or only insufficiently included in
previous simulations of $^{19}$F nucleosynthesis. The results of
all these studies will be summarized and discussed in the
following section. The main implication for the present study is
that the new experimental results put a more stringent limit on
the reaction rates and therefore reduce considerably the
associated uncertainties compared to the uncertainties listed in
the NACRE compilation \citep{angulo:99}. 
There has not been much
improvement in the $^{18}$O(p,$\alpha$)$^{15}$N rate and there has
been very little experimental effort in the study of
$^{19}$F($\alpha,p$)$^{22}$Ne. We therefore will discuss the
present nuclear physics related uncertainties associated with both
rates. For the latter case we will also give a new reaction rate
estimate based on experimental information and nuclear structure
information on the compound nucleus $^{23}$Na rather than on
simple penetrability arguments.

\subsection{The reaction rate of $^{13}$C($\alpha,n$)$^{16}$O}

For the $^{13}$C($\alpha,n$)$^{16}$O reaction, we have used the rate from 
\citet{drotleff:93} and 
\citet{denker:95} is about 50\% lower than the rate recommended by NACRE in the 
temperature range of interest. 
Recent $^{13}$C($^{6}$Li,d) $\alpha$-transfer studies \citep{kubono:03} suggest a 
very small spectroscopic factor of S$_{\alpha}$=0.01 for the subthreshold state 
at 6.356 MeV. This indicates that the high energy tail for this state is 
negligible for the reaction rate, in agreement with the present lower limit. 
However, a detailed re-analysis by \citet*{keeley:03} of the transfer data 
leads to significantly different results for the spectroscopic factor of the 
subthreshold state S$_{\alpha}$=0.2 which would imply good agreement with the 
value used in this paper. This 
situation requires further experimental and theoretical study. A re-evaluation 
of the rate based on new experimental results has 
been performed by \citet{heil:02} and will be published in a forthcoming paper. 
The choice of the \ct($\alpha,n$)\osi\ reaction within the current 
possibilities only slightly affects the production of \nfi\ and \fni. 
Using the rate by \citet{denker:95} in the 3 \msun\ $Z$=0.02 model with a 
partial mixing zone of mass 0.002 \msun\ gives 
an 8\% increase in the final surface \fni\ with respect to the calculation done 
using the NACRE rate. This result can be understood when the 
$^{13}$C($\alpha,n$)$^{16}$O rate is compared 
to the $^{14}$C($\alpha,\gamma$)$^{18}$O reaction, as discussed in the next 
subsection.

\subsection{The reaction rate of $^{14}$C($\alpha,\gamma$)$^{18}$O}

The reaction $^{14}$C($\alpha,\gamma$)$^{18}$O has been studied
experimentally in the energy range of 1.13 to 2.33 MeV near the
neutron threshold in the compound nucleus $^{18}$O by
\citet{goerres:92}. The reaction rate is dominated at higher  
temperatures by the direct capture and the single strong 4$^+$
resonance at E$_{cm}$=0.89 MeV. Toward lower temperatures, which
are of importance for He shell burning in AGB stars, important
contributions may come from the 3$^-$ resonance at E$_{cm}$=0.176
MeV (E$_x$=6.404 MeV) and a 1$^-$ subthreshold state at E$_x$=6.198
MeV. It has been shown in detailed cluster model simulations that 
neither one of the two levels is characterized by a pronounced 
$\alpha$ cluster structure \citep{descouvemont:85}. The strengths
of these two contributions are unknown and have been estimated by
\citet*{buchmann:88} adopting an $\alpha$ spectroscopic factor of
$\Theta_{\alpha}^2$=0.02, 0.06 for the 6.404 MeV and the 6.198 MeV
states, 
for determining the 0.176 MeV resonance strength and the cross
section of the high energy tail of the subthreshold state. While
the value for the 6.404 MeV state is in agreement with the results of a   
$^{14}$C($^6$Li,d)$^{18}$O $\alpha$-transfer experiment
\citep{cunsolo:81} the value for the 6.2 MeV state appears rather
large since the corresponding $\alpha$ transfer was not observed.
This reflects the lack of appreciable $\alpha$ strength in agreement  
with the theoretical predictions. We therefore adopted an upper 
limit for the spectroscopic factor of this resonance of
$\Theta_{\alpha}^2$=0.02. The upper limit for the reaction rate is
based on the experimental data \citep{goerres:92} plus the low
energy resonance contributions calculated from the upper limit for
the $\alpha$ spectroscopic factor. For the recommended reaction
rate we adopted a considerably smaller spectroscopic factor
$\Theta_{\alpha}^2$=0.01 for calculating the $\omega\gamma$
strength of the 0.176 MeV resonance. In this we followed the   
recommendations by \citet{funck:89}. The lower limit of the
reaction rate neglects the contribution of this resonance  
altogether and corresponds directly to the experimental results
\citep{goerres:92}. It should be noted however that the uncertainty
for the resonance strength and therefore its contribution to the
reaction rate is up to five orders of magnitudes as shown in 
\fig{fig:c14ago18agf19ap}.

The $^{14}$C($\alpha,\gamma$)\oei\ reaction can be activated together with
the \ct($\alpha,n$)\osi\ during the interpulse period, 
both in the partial mixing zone as well as in the deepest layer of the 
region composed by H-burning ashes, when 
\nfo($n,p$)$^{14}$C occurs, and represents the main path to the 
production of \oei, and subsequently of \nfi. The importance of the 
nucleosynthesis of \nfi\ during the interpulse periods is very much
governed by the choice of the rate of the $^{14}$C($\alpha,\gamma$)\oei\ 
reaction. The closer, or higher, this rate is to that of the 
\ct($\alpha,n$)\osi\ reaction the more efficient is the production of \nfi\ 
because \oei\ and protons are produced together. 
The effect of the partial mixing zone, and hence the uncertainties 
related to it, are in fact much less important when using our recommended rate, since 
in the temperature range of interest our rate is more than an order of magnitude lower 
than our standard rate from NETGEN \citep{jorissen:01}, which was also used in the previous 
study by \citet{goriely:00} (see \fig{fig:c13anc14ag}). At the temperature of interest the 
NETGEN rate is based on previous theoretical studies by \citet{funck:89} and 
\citet{hashimoto:86}. When using our recommended rate 
to compute the 3 \msun\ $Z=$0.02 model with a partial mixing zone of mass 
0.002 \msun, the final [\fni/\osi] is the same as that computed without the 
partial mixing 
zone within 10\%.

\subsection{The reaction rate of $^{14}$N($\alpha,\gamma$)$^{18}$F}

The low energy resonances in $^{14}$N($\alpha,\gamma$)$^{18}$F
have recently successfully been measured by \citet{gorres:00}. Previous
uncertainties about the strengths of these low energy resonances
were removed. Due to these 
results the reaction rate is reduced by about a factor of three compared 
to NACRE.

The \nfo($\alpha,\gamma$)$^{18}$F is inefficient at the
temperature of neutron release in the partial mixing zone while it
is activated in the convective pulse. Hence its rate only affects the
production of \fni\ in the pulse. Using the new rate by 
\citet{gorres:00} with respect to the rate by \citet{caughlan:88} (CF88), which is 
the same as NACRE within 10\%, only very marginally changes the 
production of \fni. For example in the 3 \msun\ $Z$=0.02 model with a 
partial mixing zone of mass 0.002 \msun\ the final 
abundance in the envelope is 
increased by about 5\% using the new rate.

\subsection{The reaction rate of $^{15}$N($\alpha,\gamma$)$^{19}$F}

The reaction rate of $^{15}$N($\alpha,\gamma$)$^{19}$F was taken  
from NACRE. The rate is dominated by the contribution
of three low energy resonances. The resonance strengths are based
on the analysis of \citet{deoliveira:96}. It should be noted
though that there were several recent experimental studies which
point towards a significantly higher reaction rate.
\citet{deoliveira:97} already suggested higher
resonance strengths than given in their earlier paper. Direct
$\alpha$-capture measurements of the two higher energy states by
\citet{wilmes:02} also indicate higher strengths. A recent     
indirect $\alpha$-transfer analysis to the three resonance levels
by \citet{fortune:03} does suggest even higher values for the
resonance strengths. Altogether the reaction rate of
$^{15}$N($\alpha,\gamma$)$^{19}$F used in this work might be
underestimated by a factor of five.

Using the reaction rate by CF88 for the \nfi($\alpha,\gamma$)\fni\, 
which is about 50 times higher with respect to the new estimate by 
\citet{deoliveira:96}, did not change the results in the 3 \msun\ $Z$=0.02 
model with a partial mixing zone of mass 0.002 \msun. The final 
\fni\ abundance in the envelope increased by few percent only. 
This is because the temperature in the thermal pulses is high enough that 
in any case all \nfi\ is transformed in \fni, as shown in 
\fig{fig:m3z02pulses}. This point was discussed by 
\citet{deoliveira:96}, who showed that at temperatures higher than 
$\simeq 2.6 \times 10^8$ K, such as those in our thermal pulses followed 
by TDU, the difference between using the two rates is minimal. 
Hence even if the final rate will actually be higher than the latest estimate,
this will not make a difference to the final results.
A maximum increase of 35\% in the final \fni\ intershell abundance would occur in 
the case of the 1 \msun\ $Z=$0.02 model, assuming that all \nfi\ 
would burn into \fni\ (see \secr{sec:allmodels}).

\subsection{The reaction rate of $^{15}$N($p,\alpha)^{12}$C}

The  $^{15}$N($p,\alpha)^{12}$C reaction has been investigated by
\citet*{schardt:52}, \citet{zyskind:79}, and more recently by
\citet{redder:82} at $E_p$(lab) = 78-810 keV.
These results were summarized and compiled by NACRE. The reaction
rate at $T_9\sim0.2$ is dominated by the $J^\pi=1^-$ resonance at
$E_p$ = 334 keV. However, contributions from three other
resonances at 1027, 1639, and 2985 keV have been included as well.
Using the NACRE rate, which is up to a 
factor of two higher than the rate by CF88, we obtain a small decrease of 
$\simeq$ 8\% in the final surface abundance of \fni\ in the 3 \msun\ $Z$=0.02 
model with a partial mixing zone of mass 0.002 \msun.

\subsection{The reaction rate of $^{18}$O($\alpha,\gamma$)$^{22}$Ne}

The \oei($\alpha,\gamma$)$^{22}$Ne is of interest for the
discussion of $^{19}$F production in AGB stars since it      
competes with the $^{18}$O(p,$\alpha$)$^{15}$N process. A strong
rate might lead to a reduction in $^{19}$F production. The
reaction rate of \oei($\alpha,\gamma$)$^{22}$Ne has been last
summarized and discussed by \citet{kappeler:94} and by the NACRE
compilation. The main uncertainties result
from the possible contributions of low energy resonances which
have been estimated on the basis of $\alpha$-transfer measurements 
by \citet{giesen:94}. A recent experimental study of
\oei($\alpha,\gamma$)$^{22}$Ne by \citet{dababneh:03} led to the  
first successful direct measurement of the postulated low energy 
resonances at 470 keV and 566 keV thus reducing to 33\% the previous 
uncertainty of about a factor of 30 given by NACRE at the temperature of 
interest which was given by taking the previously available 
experimental upper limit for the 470 keV resonance strength \citep{giesen:94}.
The new rate is shown in \fig{fig:c14ago18agf19ap}.
Not measured still is the 218 keV resonance which is expected to dominate the rate at
temperatures of T$\le$0.1 GK, well below the temperature in
typical He-burning conditions. The resulting reaction rate is in
very good agreement with the previous estimate by \citet{kappeler:94} 
which was used for our calculations of $^{19}$F production. 

\subsection{The reaction rate of $^{18}$O(p,$\alpha$)$^{15}$N}

The reaction $^{18}$O(p,$\alpha$)$^{15}$N provides a major link
for the production process of $^{19}$F. The reaction cross section
has been measured by \citet{lowi:78} down to energies of
$\approx$70 keV. Possible contributions of low energy near
threshold resonances were determined by \citet{wiescher:82} and
\citet{champagne:86} using direct capture and single particle
transfer reaction techniques. These results were compiled and
summarized by NACRE. The reaction rate uncertainties
are less than an order of magnitude, and less than a factor of two in the 
range of temperature of interest, and are mainly related to
uncertainties in the reasonably well studied single particle     
structure of these threshold resonance states.
The NACRE rate is the same within 10\% of the rate given by 
CF88. Hence we do not currently have major uncertainties on the \fni\ 
production coming from this rate. 

\subsection{The reaction rate of $^{19}$F($\alpha,p$)$^{22}$Ne}

The reaction rate of $^{19}$F($\alpha,p$)$^{22}$Ne is one of the
most important input parameters for a reliable analysis of $^{19}$F
nucleosynthesis at AGB stars. Yet, there is very 
little experimental data available for the
$^{19}$F($\alpha,p$)$^{22}$Ne reaction cross section at low  
energies. Experiments were limited to the higher energy range
above E$_{\alpha}$=1.3MeV \citep{kuperus:65}. \citet{caughlan:88}
suggested a rate which is based on a simple barrier penetration  
model previously used by \citet{wagoner:64}. This reaction rate is
in reasonable agreement with more recent Hauser-Feshbach estimates
assuming a high level density \citep[see][]{thielemann:86} and has
therefore been used in most of the previous nucleosynthesis
simulations. The applicability of the Hauser-Feshbach model,
however, depends critically on the level density in the compound
nucleus system \citep*{rauscher:97}. We analyzed the level density in the 
compound nucleus $^{23}$Na above the $\alpha$-threshold of
Q$_{\alpha}$=10.469 MeV as compiled by \citet{endt:78} and
\citet{endt:90}. The typical level density is $\approx$0.02   
keV$^{-1}$. This level density is confirmed directly for the $^{19}$F($\alpha,p$) 
reaction channel by direct studies from \citet{kuperus:65} at resonance energies 
above 1.5 MeV and further confirmed by as yet unpublished low energy 
$^{19}$F($\alpha,p$) resonance measurements of \citet{ugalde:04}. This low 
resonance density translates into an averaged level 
spacing of D$\approx$50 keV which is considerably larger than the
average resonance width of $\Gamma\approx$8 keV in this excitation
range. Based on these estimates the requirement of $D\le\Gamma$
for the applicability of the Hauser-Feshbach approach \citep{rauscher:97}
is not fulfilled. The reaction rate for
$^{19}$F($\alpha,p$)$^{22}$Ne therefore needs to be calculated 
from determining the strengths $\omega\gamma$ for the single
resonances,
\begin{equation}
\omega\gamma=\frac{(2J+1)}{2}\cdot\frac{\Gamma_{\alpha}\Gamma_p}{\Gamma_{tot}}.
\end{equation}
We estimated the $\alpha$ partial width $\Gamma_{\alpha}$ using a
simple WKB approximation with an average $\alpha$-spectroscopic
factor of C$^2$S$_{\alpha}$=0.001. This average spectroscopic
factor was determined from determining the average $\alpha$-strength 
distribution from the strengths of observed $\alpha$
capture resonances at higher energies \citep{kuperus:65} and from
the $\alpha$ spectroscopic strengths of bound states in $^{23}$Na 
\citep{fortune:78}. The total widths $\Gamma_{tot}$ of the levels
correspond in all cases to the proton partial widths $\Gamma_p$, 
therefore, the resonance strength depends entirely on the spin $J$
and the $\alpha$ partial width $\Gamma_{\alpha}$ of the resonance
levels. For the higher energy range E$_{\alpha}\ge$ 1.5 MeV we  
used directly the experimentally determined resonance strengths by
\citet{kuperus:65}. The resulting reaction rate is shown in 
\fig{fig:c14ago18agf19ap} and deviates  
considerably from the Hauser-Feshbach prediction, in the  
temperature range of intershell He burning it is more than one
order of magnitude smaller than predicted in the Hauser-Feshbach 
estimate. The possibility of ``missing strength'' in as yet unobserved 
resonances seems unlikely as shown by the previous \fni($\alpha,p)$ studies but 
cannot be completely excluded. However a substantial increase in the reaction rate 
would rather be associated with a large $\alpha$ strength of the low energy 
unbound states in $^{23}$Na. 
Therefore an experimental confirmation of the here predicted
resonance strength distribution is desirable for a wide energy 
range.

Using our new recommended rate, for example in the 3 \msun\ $Z=$0.02 model 
the final [\fni/\osi] is 0.1 dex higher than in the case computed using 
the CF88 rate. The effect of this rate and its uncertainties 
is larger for higher mass models, where the temperature is higher and 
the $^{19}$F($\alpha,p$)$^{22}$Ne is more activated.

\subsection{Other rates of interest}

The \ct($p,\gamma$)$^{14}$N reaction is of interest regarding the 
formation of \ct\ in the partial mixing zone. The experimental rate 
by \citet{king:94} is 1.29 times higher than the rate given by
CF88 at the temperature of interest, and the
revision by NACRE, which we used, gives a rate 1.20 times higher
than CF88. A higher rate will result in a lower \ct\ abundance
and a lower neutron flux during the
interpulse period. Calculations for the 3 \msun\ $Z$=0.02 model 
showed that the difference of 10\% less between NACRE and 
the rate by \citet{king:94} yields a 5\% increase in the \nfi\ 
produced during the interpulse, and a 6\% increase in the 
final surface \fni. We also checked that within the current 
uncertainties of the \nfo($n,p$)$^{14}$C rate \citep[$\simeq 10$\%,][]{gledenov:95}  
and the less important \nfo($n,\gamma$)$^{15}$N rate \citep[uncertainties of a 
factor of about 2.5,][]{beer:92}, the final results do not change.

\section{Discussion and conclusions}
\label{sec:discussion}

Using the new rates presented in the previous section, in particular for the 
$^{14}$C($\alpha,\gamma$)\oei\ and the \fni($\alpha,p$)$^{22}$Ne reactions, we have 
calculated recommended, upper and lower limits for the production of 
\fni\ in selected stellar models (\tab{tab:tab2}). The runs computed with 
no inclusion of the partial mixing zone (Column 2) can 
be considered, within our models, as absolute lower limits for the \fni\ 
yields.
The runs computed with the recommended rates and including the partial mixing zone 
(Column 4) show a decrease in the yield with respect to the same runs 
computed with the ``standard'' rates (Column 3), 
except for the 5 \msun\ $Z=$0.02 model. This decrease  
is due to our estimate of the $^{14}$C($\alpha,\gamma$)$^{18}$O reaction, which 
makes the contribution of the partial mixing zone to the production of \fni\ 
much less significant. In the case of the 5 \msun\ $Z=$0.02 model the yield 
increases of a factor of two owing to the fact that the temperatures in this 
intermediate-mass model are higher than in the other models and hence the 
effect of our lower 
estimate for the \fni($\alpha,p)^{22}$Ne rate is more important.
The overall uncertainties in the \fni\ production due to the uncertainties 
in the reaction rates are about 50\% in the stellar models with mass $\simeq$ 3 
\msun, and about 40\% in stellar models of lower mass. For the 5 
\msun\ $Z=$0.02 stellar model the uncertainties are about a factor of 5, 
due to the large uncertainties of the \fni($\alpha,p)^{22}$Ne
rate.     

The \fni($\alpha,p)^{22}$Ne reaction rate also influences the production 
of fluorine in the winds of Wolf-Rayet stars hence models of this type of 
stars should also be revised to test the effect of our revised rate 
and its uncertainties. It is also important to note that our estimated 
lower limit for the \fni($\alpha,p$)$^{22}$Ne rate is about 4 orders of 
magnitude lower than the $^{22}$Ne($\alpha,n$)$^{25}$Mg reaction rate.
In this case the \fni($n,\gamma$)$^{20}$F reaction has to be taken into 
account as a possible destruction channel for \fni\ when a significant 
neutron flux is released in the convective pulses of AGB stars and in 
Wolf-Rayet stars by the $^{22}$Ne($\alpha,n$)$^{25}$Mg reaction.  

For the 3 \msun\ $Z$=0.02 model surface abundances are also shown in
\fig{fig:f19rates} for a given choice of the partial mixing zone with 
$M_{pmz}$ = 0.002 \msun. With the new estimate for the 
$^{14}$C($\alpha,\gamma$)$^{18}$O 
rate the contribution of the partial 
mixing zone is diminished, making this uncertain parameter less 
important. In particular, in the lower limit case, the resulting [\fni/\osi] ratio 
is the same within 10\% as computed without including the partial mixing zone 
(compare to \fig{fig:f19pocket}). 
In none of the cases we calculated could the highest 
[\fni/\osi] values observed be reproduced. As discussed in 
\secr{sec:allmodels}, this problem should be reviewed with the inclusion in 
future calculations of extra-mixing processes (Cool Bottom Processing) at the 
base of the convective envelope.

Future work should also improve our knowledge of the formation and the 
nucleosynthesis 
in the partial mixing zone. One hypothesis is that rotation can play a role in 
varying the efficiency of the production of \fni\ and
of the \spr\ elements \citep{herwig:03}. It will be of much interest to analyze the
effects of this hypothesis on the correlation between fluorine and the 
$s$-process elements and to revise the available observational data. Using data for 
carbon stars from \citet{utsumi:85} it appeared that these two quantities were 
correlated in AGB stars, however using more recent and precise data from 
\citet{abia:02} this correlation does not seem to appear anymore. 

Another problem is related to C(J) stars. It is still unknown if these 
stars actually belong to the AGB group or if they are in some other
phase of the evolution. Moreover, it appears that their
[\fni/\osi] ratios around 0.6 are due to a low abundance of \osi\
rather than a high abundance of \fni. Finally, the observational
data regarding SC stars should be
updated using more recent atmospheric models.

\acknowledgments
M. L. deeply appreciates the hospitality and support extended to her by 
Michael Wiescher, Joachim G\"orres and Claudio Ugalde during a visit to Notre 
Dame  University, as well as the
hospitality received by John Lattanzio and Amanda Karakas during a visit 
to Monash University.
We thank Roberto Gallino, Enrico Arnone, Stefano Masera and Richard 
Stancliffe for discussion and help.
The manuscript was much extended and improved following strong criticisms 
by the anonymous referee. This work was supported by the Australian Research 
Council and the Australian Partnership for Advances Computing. 
Computational resources used for this study were also partly funded by the
Canada Foundation for Innovation (CFI) and the Nova Scotia Research and   
Innovation Trust fund (NSRIT).

\newpage

\appendix
\section{Details of the reaction rates used in the reference case}
\label{app:rates}

References for proton, $\alpha$ and neutron captures that we have used in the 
nucleosynthesis calculations are presented in \tab{tab:pcaptures}, \tab{tab:acaptures}
and \tab{tab:ncaptures} respectively. All the reaction not listed in the tables are 
taken from the REACLIB Data Tables (version 1991).

\clearpage

\begin{deluxetable}{ccccc}
\tablecolumns{5}
\tablewidth{0pc}
\tablecaption{\label{tab:tab1}
$^{19}$F yields in solar masses from all the computed stellar models$^a$}
\tablehead{\colhead{$M$(\msun), $Z$} & \colhead{0.02} & \colhead{0.008} & 
\colhead{0.004} & \colhead{0.0001}}
\startdata

1. & 3.65($-$8) & 2.37($-$9) & 9.45($-$10) & 5.14($-$9) \\
1.25 & 1.59($-$8) & 1.14($-$8) & 6.23($-$9) & 2.12($-$7) \\
1.50 & 2.51($-$8) & 2.02($-$8) & 2.29($-$8) & 5.43($-$6) \\
1.75 & 3.01($-$8) & 9.01($-$8) & 1.73($-$7) & \\
1.90 & 2.83($-$8) & 1.87($-$7) & 4.96($-$7) & \\
2.00 & 2.72($-$8) & 6.41($-$7) & & 1.05($-$5) \\ 
2.25 & 1.20($-$7) & 1.49($-$6) & 3.87($-$6) & 1.36($-$5) \\
2.50 & 9.95($-$7) & 3.36($-$6) & 8.10($-$6)  & 4.56($-$6) \\
3.00 & 3.93($-$6) & 9.98($-$6) & 6.89($-$6)  & 6.20($-$8) \\
3.50 & 6.00($-$6) & 2.52($-$6) & 8.17($-$7) & \\
4.00 & 2.07($-$6) & 8.33($-$7) & 8.90($-$8) & 2.74($-$9) \\
5.00 & 6.12($-$7) & $-$1.18($-$6) & $-$6.50($-$7) & $-$6.94($-$9) \\
6.00 & $-$2.18($-$6) & $-$1.62($-$6) & $-$8.41($-$7) & \\
6.50 & $-$2.45($-$6) & &  \\

\enddata

\vskip 0.5 cm
{\footnotesize $^a$As in \fig{fig:yields}: no partial mixing zone included and 
reaction rates from Appendix \ref{app:rates}.}

\end{deluxetable}

\clearpage

\begin{landscape}

\begin{deluxetable}{cccccc}
\tablecolumns{6}
\tablewidth{0pc}
\tablecaption{\label{tab:tab2}
$^{19}$F yields in solar masses from selected stellar models$^a$}
\tablehead{\colhead{Rates:} & \colhead{standard$^b$} & \colhead{standard$^b$} & 
\colhead{recommended$^c$} & \colhead{upper$^d$} & \colhead{lower$^e$}}
\startdata

$M_{pmz}$(\msun) = & 0 & 0.002 & 0.002 & 0.002 & 0.002 \\

3, 0.02 & 3.93($-$6) & 1.01($-$5) & 6.49($-$6) & 7.10($-$6) & 4.78($-$6) \\
5, 0.02 & 6.12($-$7) & 2.37($-$6) & 4.35($-$6) & 5.18($-$6) & 9.60($-$7) \\
1.75, 0.008 & 9.01($-$8) & 1.23($-$6) & 5.39($-$7) & 5.96($-$7) & 4.37($-$7) \\
3, 0.008 & 9.98($-$6) & 2.23($-$5) & 1.94($-$5) & 2.09($-$5) & 1.36($-$5) \\
2.5, 0.004 & 8.10($-$6) & 1.96($-$5) & 1.60($-$5) & 1.69($-$5) & 1.21($-$5) \\

\enddata

\vskip 0.5 cm

{\footnotesize $^a$Since each model run takes at least one CPU day 
it is unfeasible to repeat all the calculations presented in 
\secr{sec:allmodels}. More calculations will be performed under specific 
requests.}
{\footnotesize $^b$As listed in Appendix \ref{app:rates}.}
{\footnotesize $^c$As described in \secr{sec:summa}, specifically: 
$^{14}$N($\alpha,\gamma)^{18}$F from \citet{gorres:00}, $^{18}$O($\alpha,\gamma)^{22}$Ne 
from \citet{dababneh:03}, and our recommended values for $^{14}$C($\alpha,\gamma)^{18}$O 
and $^{19}$F($\alpha,p)^{22}$Ne.}
{\footnotesize $^d$Upper limit for the $^{14}$C($\alpha,\gamma)^{18}$O rate 
and 
lower limit for the $^{19}$F($\alpha,p)^{22}$Ne rate (\secr{sec:summa}) to obtain the 
upper limit for the yields.}
{\footnotesize $^e$Lower limit for the $^{14}$C($\alpha,\gamma)^{18}$O rate 
and 
upper limit for the $^{19}$F($\alpha,p)^{22}$Ne rate (\secr{sec:summa}) to obtain the 
lower limit for the yields.}

\end{deluxetable}
\end{landscape}

\begin{deluxetable}{cc}
\tablecolumns{2}
\tablewidth{0pc}
\tablecaption{\label{tab:pcaptures}
Proton captures}
\tablehead{\colhead{reaction} & \colhead{reference}}
\startdata

$^{7}$Be(p,${\gamma}$)$^{8}$B & \citet{hammache:98} \\

$^{13}$C(p,${\gamma}$)$^{14}$N & NACRE \citep{angulo:99} \\

$^{14}$C(p,${\gamma}$)$^{15}$N & \citet*{wiescher:90} \\

$^{13}$N(p,${\gamma}$)$^{14}$O & \citet{decrock:93} \\

$^{17}$O(p,${\gamma}$)$^{18}$F & \citet{blackmon:95,landre:90} \\

$^{17}$O(p,${\alpha}$)$^{14}$N & \citet{blackmon:95,landre:90} \\

$^{18}$F(p,${\gamma}$)$^{19}$Ne & \citet{utku:98} \\

$^{18}$F(p,${\alpha}$)$^{15}$O & \citet{utku:98} \\

$^{21}$Ne(p,${\gamma}$)$^{22}$Na & \citet{eleid:95} \\

$^{22}$Ne(p,${\gamma}$)$^{23}$Na & \citet{eleid:95} \\

$^{22}$Na(p,${\gamma}$)$^{23}$Mg & \citet{stegmuller:96,schmidt:95,seuthe:90} \\

$^{23}$Na(p,${\gamma}$)$^{24}$Mg & \citet{eleid:95} \\

$^{23}$Na(p,${\alpha}$)$^{20}$Ne & \citet{eleid:95} \\

$^{24}$Mg(p,${\gamma}$)$^{25}$Al & \citet{powell:99} \\

$^{25}$Mg(p,${\gamma}$)$^{26}$Al$^{g/i}$ & \citet{iliadis:96,iliadis:90}\\

$^{26}$Mg(p,${\gamma}$)$^{27}$Al & \citet{iliadis:90} \\

$^{26}$Al$^{g}$(p,${\gamma}$)$^{27}$Si & \citet{champagne:93,vogelaar:96} \\

$^{27}$Al(p,${\gamma}$)$^{28}$Si & \citet{iliadis:90,timmermann:88} \\

$^{27}$Al(p,${\alpha}$)$^{24}$Mg & \citet{timmermann:88,champagne:88} \\

$^{28}$Si(p,${\gamma}$)$^{29}$P & \citet{graff:90} \\

\enddata

\end{deluxetable}

\clearpage
\begin{deluxetable}{cc}
\tablecolumns{2}
\tablewidth{0pc}
\tablecaption{\label{tab:acaptures}
$\alpha$ captures}
\tablehead{\colhead{reaction} & \colhead{reference}}
\startdata

$^{13}$C(${\alpha}$,n)$^{16}$O & \citet{drotleff:93,denker:95} \\

$^{14}$C(${\alpha}$,${\gamma}$)$^{18}$O & NETGEN \citep{jorissen:01} \\

$^{15}$N(${\alpha}$,${\gamma}$)$^{19}$F & \citet{deoliveira:96} \\

$^{17}$O(${\alpha}$,n)$^{20}$Ne & \citet{denker:95} \\

$^{18}$O(${\alpha}$,${\gamma}$)$^{22}$Ne & \citet{kappeler:94,giesen:94} \\

$^{18}$O(${\alpha}$,n)$^{21}$Ne & \citet{denker:95} \\

$^{21}$Ne(${\alpha}$,n)$^{24}$Mg & \citet{denker:95} \\

$^{22}$Ne(${\alpha}$,${\gamma}$)$^{26}$Mg$^{a}$ & \citet{kappeler:94} \\

$^{22}$Ne(${\alpha}$,n)$^{25}$Mg$^{a}$ & \citet{kappeler:94,drotleff:93} \\

\enddata
\flushleft{$^{a}$The elusive resonance at 633 keV has not been included.}

\end{deluxetable}

\begin{deluxetable}{cc}
\tablecolumns{2}
\tablewidth{0pc}
\tablecaption{\label{tab:ncaptures}
Neutron captures}
\tablehead{\colhead{reaction$^{a}$} & \colhead{reference}}
\startdata

$^{12}$C(n,${\gamma}$)$^{13}$C & \citet{kikuchi:98} \\

$^{13}$C(n,${\gamma}$)$^{14}$C & \citet{raman:90}; A. Mengoni (1998, private 
communication) \\

$^{14}$N(n,p)$^{14}$C & \citet{gledenov:95} \\

$^{16}$O(n,${\gamma}$)$^{17}$O & \citet{igashira:95} \\

$^{18}$O(n,${\gamma}$)$^{19}$O & \citet{meissner:96} \\

$^{26}$Al$^{g}$(n,p)$^{26}$Mg & \citet{koehler:97} \\

$^{26}$Al$^{g}$(n,$\alpha$)$^{26}$Mg & \citet{koehler:97}; \citet*{skelton:87} \\

$^{33}$S(n,${\alpha}$)$^{30}$Si & \citet{schatz:95} \\

\enddata

\flushleft{$^{a}$The ($n,\gamma$) reactions on stable nuclei
not listed here are all from \citet*{beer:92}. Among those, the $^{28}$Si ($n,\gamma$) 
cross section has been renormalized to the value given by \citet{bao:87} 
following H. 
Beer (1990, private communication).} 

\end{deluxetable}

\clearpage
\centerline{\bf FIGURE CAPTIONS}

\figcaption[fig1]{\label{fig:intershell}
Mass fraction of \fni\ in the He intershell after the last thermal   
pulse computed for each model. No partial mixing zone was included in
these calculations.}

\figcaption[fig2]{\label{fig:yields}
Yield of \fni\ for each of the models presented in \fig{fig:intershell}.}

\figcaption[fig3]{\label{fig:f19models}
Comparison of fluorine abundances observed by \citet{jorissen:92} and model
predictions for selected stellar models: 3 and 5 \msun\ with $Z$=0.02;
1.75 and 3 \msun\ with $Z$=0.008; and 2.5 \msun\ with $Z$=0.004. 
Predictions are normalized in such way that the initial \fni\ abundance 
corresponds to the 
average F abundance observed in K and M stars, to which stellar data are 
normalized \citep[see][]{jorissen:92}. Each symbol on the prediction lines 
represents a TDU 
episode. Note that for the 2.5 \msun\ $Z$=0.004 model the final C/O=11 
and [\fni/\osi]=1.7 are outside the range of the plot. Crossed MS, S symbols 
denote stars with large N excesses.
}

\figcaption[fig4]{
\label{fig:m3z02pulses}
Abundance in number of \nfi\ (crosses) and \fni\ (full dots) in the He 
intershell as function of the pulse number for the 3 \msun\ $Z$=0.02 model with a 
partial mixing zone of mass 0.001 
\msun\ included after each TDU episode, i.e. after the 10$^{\rm th}$ thermal 
pulse. Abundances are plotted only during the time when
the convective shell is present. The final abundances for each pulse are  
those corresponding to the pulse number tick-mark in the x-axis. 
}

\figcaption[fig5]{
\label{fig:f19pocket}
Comparison of fluorine abundances observed by \citet{jorissen:92} and model
predictions for the 3 \msun\ $Z$=0.02 model and different choices of the 
extent of the partial mixing zone. The bare line represents the cases in which 
no partial mixing zone is included. The lines accompanied by tiny full or open 
dots refer to cases computed
with a partial mixing zone with mass extent $M_{pmz}$=0.001 and 0.002 
\msun, respectively.
As in \fig{fig:f19models} crossed MS, S symbols denote stars with 
large N excesses and predictions 
are normalized in such way that the initial \fni\ abundance corresponds to the 
average F abundance observed in K and M stars, to which stellar data are 
normalized \citep[see][]{jorissen:92}. 
}

\figcaption[fig6]{
\label{fig:c14ago18agf19ap} Recommended, lower and upper limits for 
the rates of the $^{14}$C($\alpha,\gamma$)$^{18}$O, 
$^{18}$O($\alpha,\gamma$)$^{22}$Ne and $^{19}$F($\alpha,p$)$^{22}$Ne
reactions. 
}

\figcaption[fig7]{
\label{fig:c13anc14ag} The rate for the $^{13}$C($\alpha,n$)$^{16}$O
reaction \citep{drotleff:93,denker:95} (solid line) is compared to two different choices 
for the $^{14}$C($\alpha,\gamma$)$^{18}$O reaction rate: NETGEN (dash-dotted 
line) and our recommended rate (dashed line) in the range of temperature at which 
the $^{13}$C($\alpha,n$)$^{16}$O reaction is activated in the $^{13}$C pocket.
}

\figcaption[fig8]{
\label{fig:f19rates}
Comparison of fluorine abundances observed by \citet{jorissen:92} and model
predictions for the 3 \msun\ $Z$=0.02 model and $M_{pmz}$=0.002 \msun\ 
and different choices of the rate of the reactions involved as described 
in \tab{tab:tab2}: ``standard'' (solid line), ``recommended'' 
(short-dashed line), lower and upper limit (dotted
lines).
As in \fig{fig:f19models} crossed MS, S symbols denote stars with large N excesses 
and predictions are normalized in such way that the initial \fni\ abundance 
corresponds to the 
average F abundance observed in K and M stars, to which stellar data are 
normalized \citep[see][]{jorissen:92}. 
}

\clearpage

\begin{figure}
\plotone{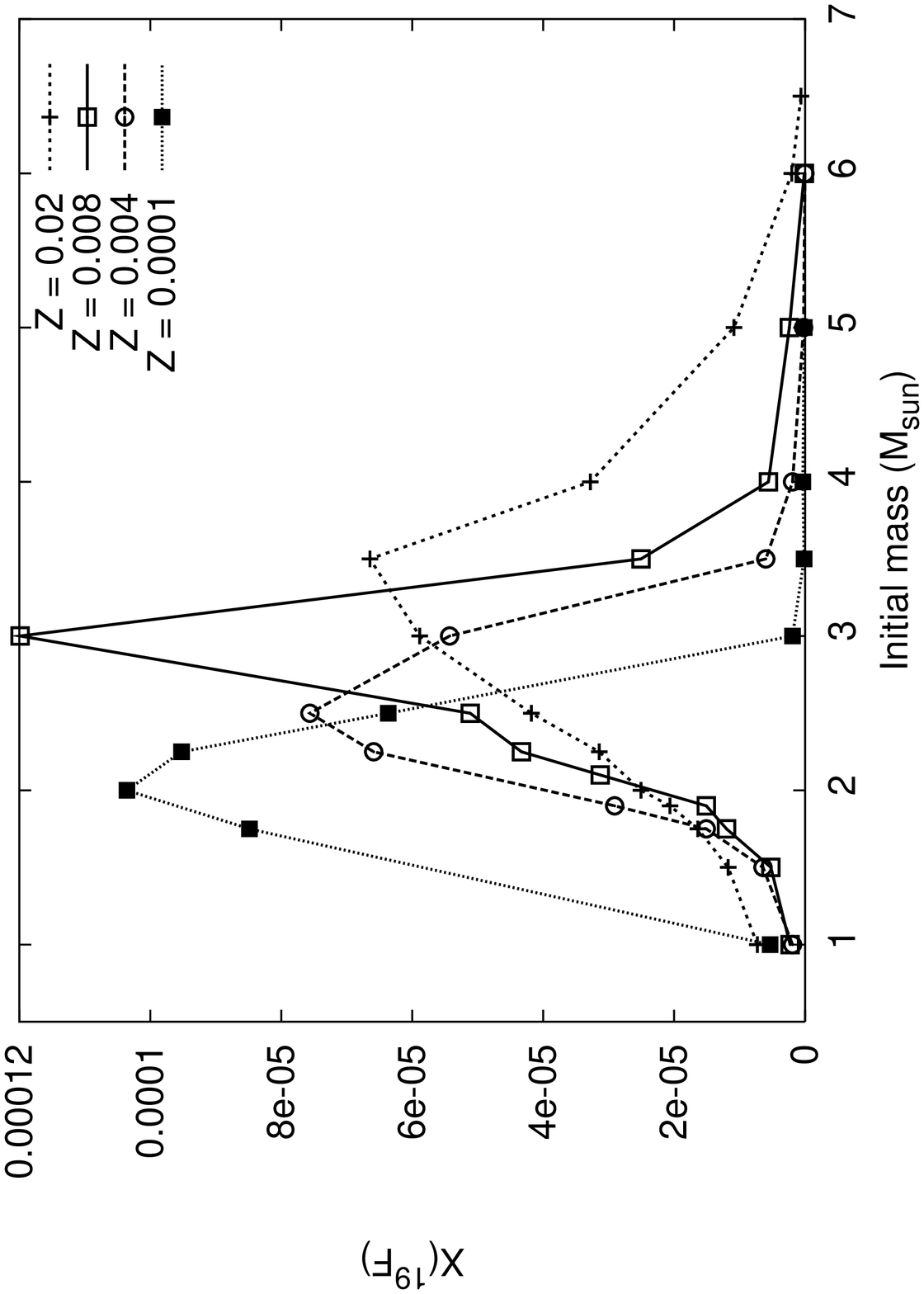}
\end{figure}

\begin{figure}
\plotone{f2.ps}
\end{figure}

\begin{figure}
\plotone{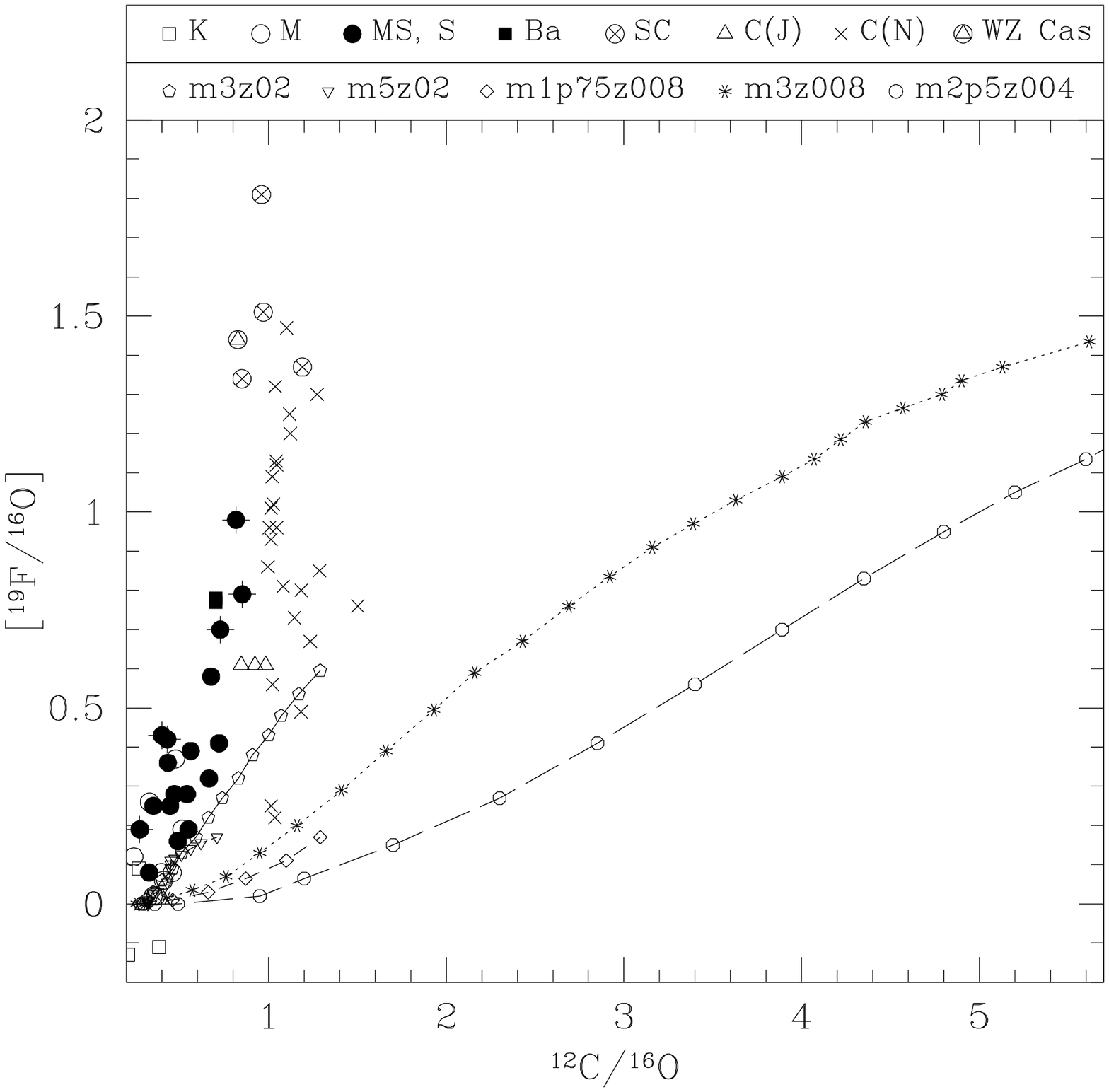}
\end{figure}

\begin{figure}
\plotone{f4.ps}
\end{figure}

\begin{figure}
\plotone{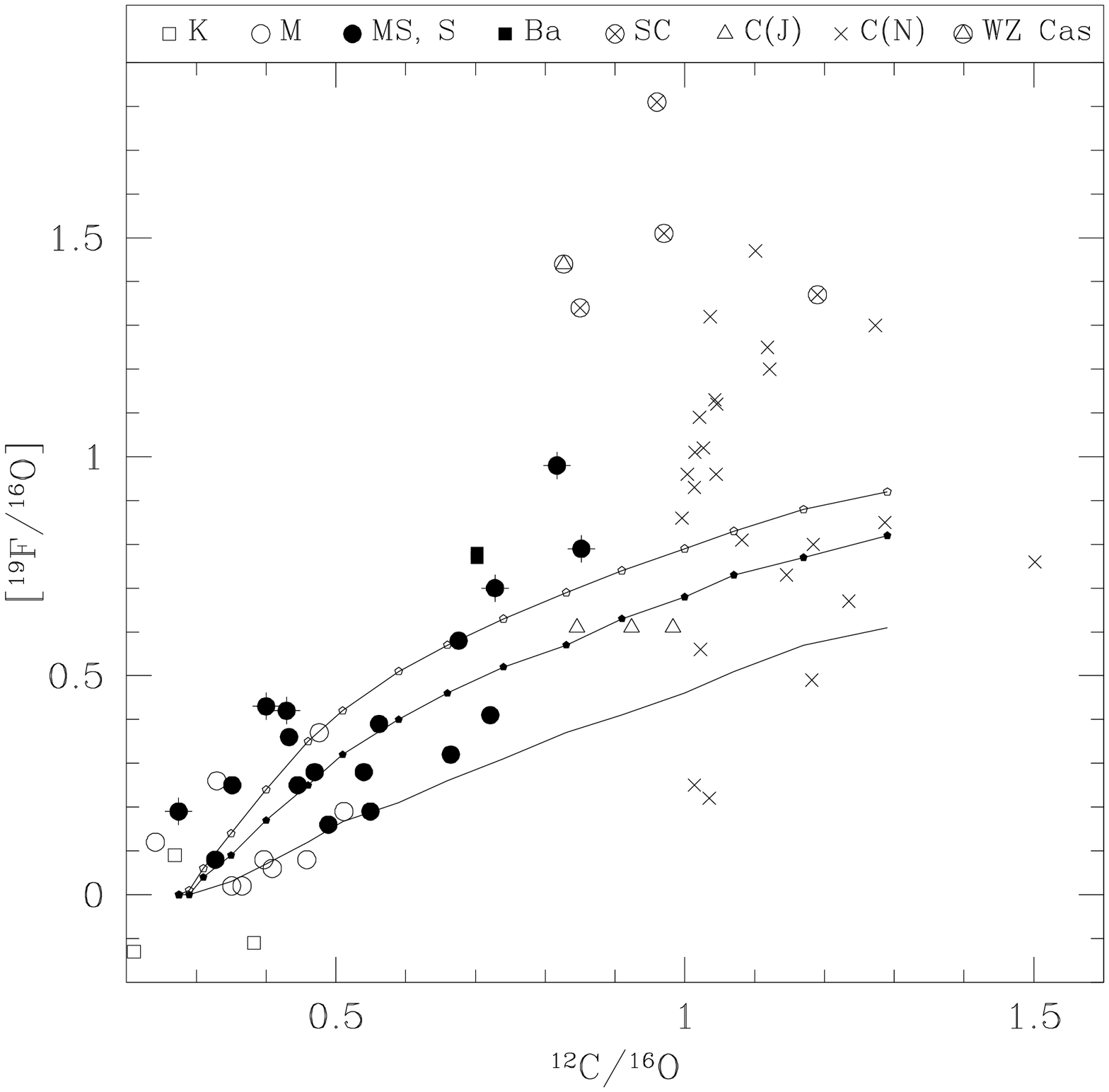}
\end{figure}

\begin{figure} 
\plotone{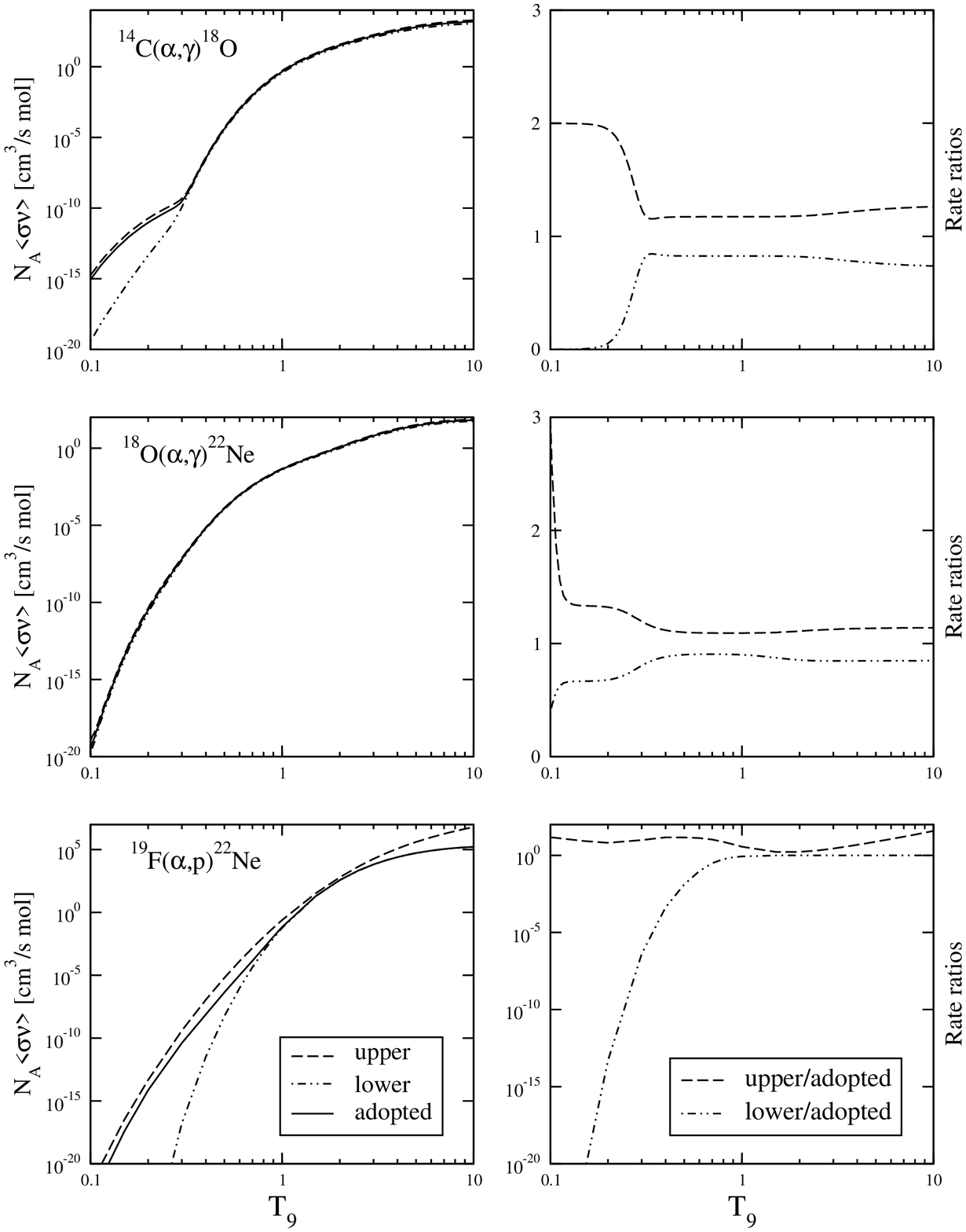} 
\end{figure}

\begin{figure} 
\plotone{f7.ps} 
\end{figure}

\begin{figure}
\plotone{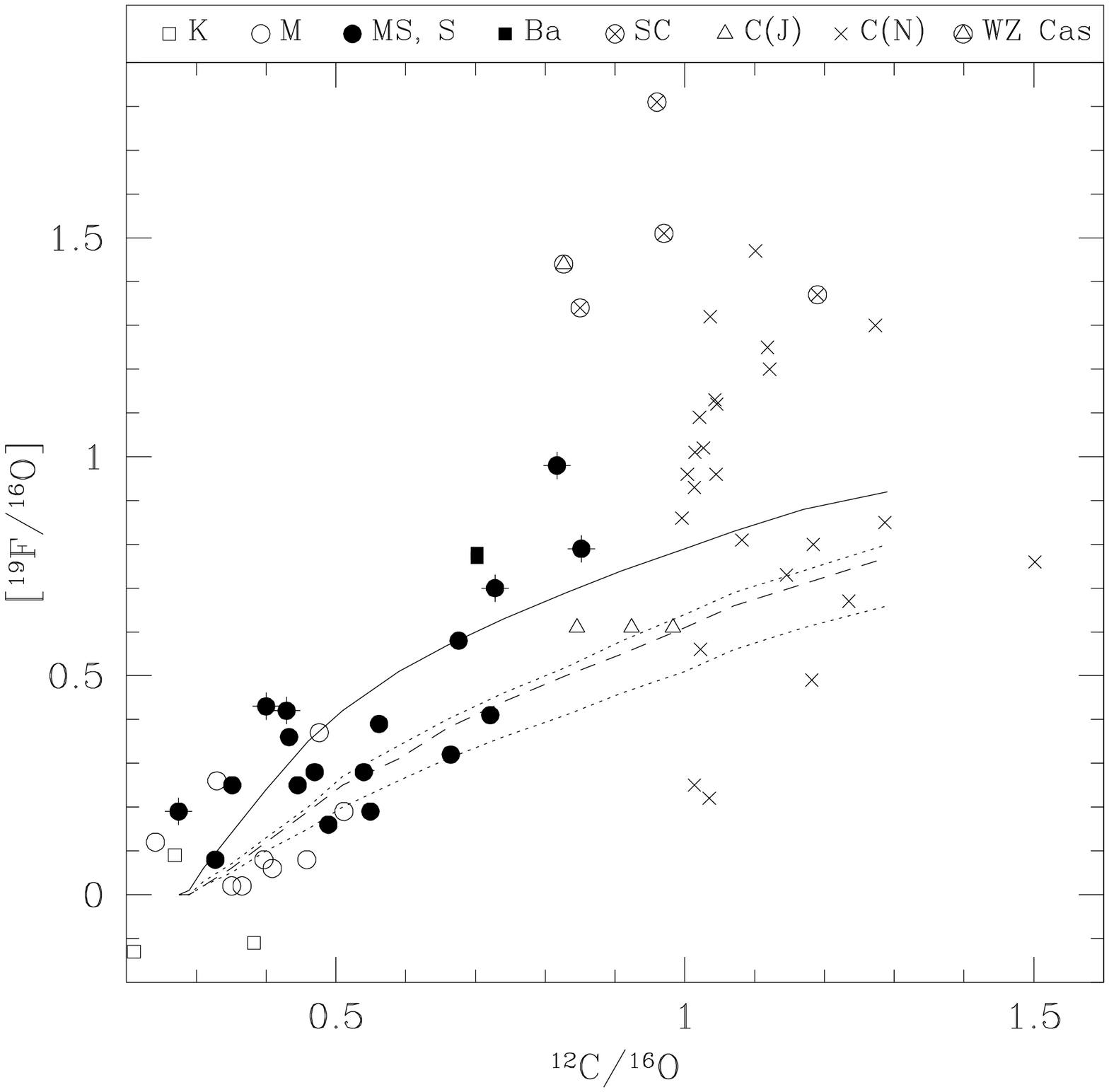}
\end{figure}

\end{document}